\documentclass[sigplan,authorversion]{acmart}

\copyrightyear{2019} 
\acmYear{2019} 
\setcopyright{acmlicensed}
\acmConference[FTfJP'19]{Formal Techniques for Java-like Programs}{July 15, 2019}{London, United Kingdom}
\acmBooktitle{Formal Techniques for Java-like Programs (FTfJP'19), July 15, 2019, London, United Kingdom}
\acmPrice{15.00}
\acmDOI{10.1145/3340672.3341119}
\acmISBN{978-1-4503-6864-3/19/07}

\usepackage{booktabs}   %% For formal tables:
\usepackage{subcaption} %% For complex figures with subfigures/subcaptions
\begin{document}

%% Title information
\title{Analysis of MiniJava Programs via Translation to ML}         %% [Short Title] is optional;
                                        %% when present, will be used in
                                        %% header instead of Full Title.
%\titlenote{with title note}             %% \titlenote is optional;
                                        %% can be repeated if necessary;
                                        %% contents suppressed with 'anonymous'
%\subtitle{Subtitle}                     %% \subtitle is optional
%\subtitlenote{with subtitle note}       %% \subtitlenote is optional;
                                        %% can be repeated if necessary;
                                        %% contents suppressed with 'anonymous'

%% Author information
%% Contents and number of authors suppressed with 'anonymous'.
%% Each author should be introduced by \author, followed by
%% \authornote (optional), \orcid (optional), \affiliation, and
%% \email.
%% An author may have multiple affiliations and/or emails; repeat the
%% appropriate command.
%% Many elements are not rendered, but should be provided for metadata
%% extraction tools.

%% Author with single affiliation.
\author{Martin Mariusz Lester}
%\authornote{with author1 note}          %% \authornote is optional;
                                        %% can be repeated if necessary
\orcid{0000-0002-2323-1771}             %% \orcid is optional
\affiliation{
%  \position{Lecturer in Computer Science}
  \department{Department of Computer Science}              %% \department is recommended
  \institution{University of Reading}            %% \institution is required
%  \streetaddress{Street1 Address1}
%  \city{City1}
%  \state{State1}
%  \postcode{Post-Code1}
  \country{United Kingdom}                    %% \country is recommended
}
\email{m.lester@reading.ac.uk}          %% \email is recommended

%% Abstract
%% Note: \begin{abstract}...\end{abstract} environment must come
%% before \maketitle command
\begin{abstract}
MiniJava is a subset of the object-oriented programming language Java.
Standard ML is the canonical representative of the ML family of functional programming languages,
which includes F\# and OCaml.
Different program analysis and verification tools and techniques have been developed
for both Java-like and ML-like languages.
Naturally, the tools developed for a particular language emphasise
accurate treatment of language features commonly used in that language.
In Java, this means objects with mutable properties and dynamic method dispatch.
In ML, this means higher order functions and algebraic datatypes with pattern matching.

We propose to translate programs from one language into the other
and use the target language's tools for analysis and verification.
By doing so, we hope to identify areas for improvement in the target language's tools
and suggest techniques, perhaps as used in the source language's tools, that may guide
their improvement.
More generally, we hope to develop tools for reasoning about programs
that are more resilient to changes in the style of code and representation of data.
We begin our programme by outlining a translation from MiniJava to ML
that uses only the core features of ML;
in particular, it avoids the use of ML's mutable references.
\end{abstract}

%% 2012 ACM Computing Classification System (CSS) concepts
%% Generate at 'http://dl.acm.org/ccs/ccs.cfm'.
\begin{CCSXML}
<ccs2012>
<concept>
<concept_id>10011007.10010940.10010992.10010998</concept_id>
<concept_desc>Software and its engineering~Formal methods</concept_desc>
<concept_significance>300</concept_significance>
</concept>
<concept>
<concept_id>10011007.10011006.10011008.10011009.10011011</concept_id>
<concept_desc>Software and its engineering~Object oriented languages</concept_desc>
<concept_significance>300</concept_significance>
</concept>
<concept>
<concept_id>10011007.10011006.10011008.10011009.10011012</concept_id>
<concept_desc>Software and its engineering~Functional languages</concept_desc>
<concept_significance>300</concept_significance>
</concept>
</ccs2012>
\end{CCSXML}

\ccsdesc[300]{Software and its engineering~Formal methods}
\ccsdesc[300]{Software and its engineering~Object oriented languages}
\ccsdesc[300]{Software and its engineering~Functional languages}
%% End of generated code

%% Keywords
%% comma separated list
\keywords{Java, ML, automated verification, static analysis, program transformation}  %% \keywords are mandatory in final camera-ready submission

%% \maketitle
%% Note: \maketitle command must come after title commands, author
%% commands, abstract environment, Computing Classification System
%% environment and commands, and keywords command.
\maketitle

\section{Motivation}

Tools for program analysis and verification have developed rapidly since the
success of Microsoft's SLAM driver verification project~\cite{DBLP:conf/ifm/BallCLR04}.
A range of complementary and overlapping techniques and technologies have gained prominence,
such as abstract interpretation, model-checking, CEGAR and SMT solvers.
% cite techs?
All provide some way of bounding potentially infinite behaviours in a program
or avoiding state space explosion.

Many of the biggest successes have been in the world of traditional imperative programs.
%for example those written in C.
Idiomatic C programs make \emph{comparatively} little use of dynamic memory allocation,
but may control their behaviour through intricate use of bit-level manipulation
and values of complex combinations of flags and other variables.
Bounded model-checking using SMT solvers
has been particularly successful
%for this kind of program
here~\cite{DBLP:conf/tacas/ClarkeKL04}.

There has also been some success in dealing with object-oriented programs,
such as those written in Java~\cite{DBLP:journals/sigsoft/CordeiroKS18},
and functional programs~\cite{DBLP:journals/dagstuhl-reports/GaboardiJJW16},
written in ML or Haskell.
%In particular, a Java track has recently been added to the SV-COMP
%software verification competition~\cite{DBLP:journals/sigsoft/CordeiroKS18},
%identifying JBMC, SPF, JayHorn and JPF as some of the leading tools.

The challenges for handling idiomatic programs written in these paradigms are different.
In Java, allocation of objects on the heap is very common.
Use of dynamic method dispatch is central to writing idiomatic Java code of any complexity.
This means that, even for simple programs,
accurate modelling of program control flow requires good modelling of the heap,
combined with context sensitivity to match method calls and returns.
(In C programs, the equivalent problem of tracking function pointers
stored at heap-allocated memory locations still arises, but less frequently.)
However, this may not always be important for program verification,
as in a well-designed object-oriented program
(or at least one that obeys the Liskov Substitution Principle),
methods of a subclass that override methods in the superclass
will usually satisfy a stronger specification than the method they override.
Thus, for many verification problems,
it is not necessary to know exactly which subclass method is being called.

In functional languages, the use of higher order functions is similarly prevalent.
Conceptually, the difficulty they present is similar to dynamic method dispatch in object-oriented programming,
but the complexity of analysis required is often greater.
Firstly, accurately tracking flow control for higher order functions
requires tracking of more levels of calling context.
Secondly, the same functionals are often used in a wide variety of unrelated situations,
so type information cannot reliably be used to delineate and partition their uses.
Furthermore, for the same reason, determining the actual results of functionals
is more important for accurate program analysis.
Consequently, many analyses for functional programming languages
emphasise accurate modelling of control flow for higher order functions.
In contrast, they often neglect or ignore mutable state,
as its use is prohibited in Haskell (other than through monads) and discouraged in ML.

\section{Goals}

Because Java and ML have different feature sets,
it is difficult to apply an analysis designed for one language to a program written in the other.
But by doing so, we may gain some insight into our tools and techniques.
We may discover that techniques developed in one community would be useful to the other.
Or the inability of one community's tool to handle programs from the other
may motivate improvements to the tool.
In particular, the introduction of lambda expressions to Java 8
may make it more important for Java tools to be able to reason about
higher order functions in programs written in a functional style~\cite{DBLP:conf/issta/Cok18}.

We propose to begin this exploration by translating Java programs into ML,
so that they may be analysed by tools written for functional programs.
In order for the translation to be manageable, we focus on translating the MiniJava subset of Java.
So that our translated programs may be used with as many tools as possible,
we use only the core features of the language,
namely recursive functions, algebraic datatypes (including lists) and pattern-matching.
In particular, we avoid the use of references (mutable variables).
Subject to these constraints, we aim to be as idiomatic as reasonably possible in our translation.

MiniJava is a subset of Java introduced in Appel and Palsberg's book
\emph{Modern Compiler Implementation in Java}\cite{DBLP:books/cu/Appel2002}.
Types in MiniJava are limited to \texttt{int}, \texttt{boolean}, arrays of \texttt{int} and
object types corresponding to any classes defined in the program.
Java features omitted from MiniJava include
interfaces, explicit casts, exceptions, visibility modifiers, generics and reflection.
%There is no input and output is limited to printing integers to standard output.
%Methods can be overridden in subclasses, but not overloaded.
%Flow control is limited to \texttt{while} loops and \texttt{if} statements.
%There are no explicit casts, but implicit upcasts are permitted.
%There are no exceptions, no generics and no visibility modifiers.
%There is no reflection and no use of \texttt{static} except for a single
%\texttt{main} method.
The combination of features is expressive enough for writing
idiomatic object-oriented programs,
but constrained enough to support easy compilation, analysis or transformation.

\section{Translation}

\paragraph{Statements and expressions.}
Each Java statement becomes a let-binding,
with the ``current'' program state being used in the bound expression
and the ``next'' program state being the newly bound variables.
The style of the resulting code is similar to
Administrative Normal Form~\cite{DBLP:conf/pldi/1993}.

\paragraph{Mutable state.}
The mutable state of a Java program is split into two parts:
heap-allocated objects and method-local variables.
As the number of local variables in any method is fixed,
the local variables can be encoded as a fixed-size tuple of variable values.
The heap is a map from pointers to objects.
Pointers can be encoded using any datatype that supports the operations required for a name,
namely comparison for equality and creation of fresh names.
The simplest choice is to use unbounded integers starting at 0,
allocating integers sequentially as fresh pointers.
Any encoding of maps can be used, but the choice will impact the analysis of the translated program.

\paragraph{Objects and subclasses.}
Java objects are encodable as a tuple combining their methods (which become ML functions)
and their properties
(which become either \texttt{int}s, \texttt{bool}s or \texttt{int}s encoding object pointers).
Member lookup simply becomes selection of an element from the tuple.
Property update requires replacing the whole object in the map encoding the heap.
Subclassing could be handled using row-level polymorphism for
records~\cite{DBLP:conf/lics/Wand89},
as in OCaml's objects.
As this is not part of Standard ML,
we instead encode an object as a tuple combining its members \emph{and}
an Option for any subclass members.
The type of the Option is then an algebraic sum over all possible subclasses.

\section{Related Work}

Program transformation
is often used for removal of more complex features of a language~\cite{DBLP:conf/popl/ChoiAYT11},
or translation to a simpler language,
so that the verification tools need only handle a smaller number of language features.
Notably, the Jimple~\cite{DBLP:conf/cascon/Vallee-RaiCGHLS99} intermediate language for Java
used by Soot is deliberately simpler than Java bytecode.
Such transformations are often avoided,
as they hide the structure of a program, confounding analysis.
Indeed, attempting to recover this structure
is a key step in analysis of compiled programs~\cite{DBLP:conf/cc/FedericoPA17}.

Previous work considers analysis of functional programs written in Haskell
via translation to C using the compiler JHC
and application of the symbolic execution tool Klee~\cite{mariopicallo}.
We are not aware of any work in the reverse direction,
presumably because of the relative immaturity of tools for functional languages.
Tools for analysing ML programs are based around a variety of different techniques,
such as model-checking of Higher Order Recursion Schemes (MoCHi~\cite{DBLP:conf/pepm/SatoUK13}),
refinement type inference (DSolve~\cite{DBLP:conf/cav/KawaguchiRJ10})
and algorithmic game semantics (SyTeCi~\cite{jabersyteci}),
but there is no clear leader.

\section{Status and Future Work}

We are currently implementing the translation.
The starting point for our work is a toy MiniJava compiler used to teach a
module on compilers at the University of Reading.
The next step will be to compare Java analysis tools on MiniJava programs
with ML program tools on the translated programs.

We expect that they will be reasonably accurate
until they have to reason about values retrieved from the heap,
however we choose to encode it.

%% Acknowledgments
%\begin{acks}                            %% acks environment is optional
%                                        %% contents suppressed with 'anonymous'
%  %% Commands \grantsponsor{<sponsorID>}{<name>}{<url>} and
%  %% \grantnum[<url>]{<sponsorID>}{<number>} should be used to
%  %% acknowledge financial support and will be used by metadata
%  %% extraction tools.
%  This material is based upon work supported by the
%  \grantsponsor{GS100000001}{National Science
%    Foundation}{http://dx.doi.org/10.13039/100000001} under Grant
%  No.~\grantnum{GS100000001}{nnnnnnn} and Grant
%  No.~\grantnum{GS100000001}{mmmmmmm}.  Any opinions, findings, and
%  conclusions or recommendations expressed in this material are those
%  of the author and do not necessarily reflect the views of the
%  National Science Foundation.
%\end{acks}

%% Bibliography
\bibliography{refs}


\begin{thebibliography}{15}

%%% the Web address.

\ifx \showCODEN    \undefined \def \showCODEN     #1{\unskip}     \fi
\ifx \showDOI      \undefined \def \showDOI       #1{#1}\fi
\ifx \showISBNx    \undefined \def \showISBNx     #1{\unskip}     \fi
\ifx \showISBNxiii \undefined \def \showISBNxiii  #1{\unskip}     \fi
\ifx \showISSN     \undefined \def \showISSN      #1{\unskip}     \fi
\ifx \showLCCN     \undefined \def \showLCCN      #1{\unskip}     \fi
\ifx \shownote     \undefined \def \shownote      #1{#1}          \fi
\ifx \showarticletitle \undefined \def \showarticletitle #1{#1}   \fi
\ifx \showURL      \undefined \def \showURL       {\relax}        \fi
% invisible to TeX
\providecommand\bibfield[2]{#2}
\providecommand\bibinfo[2]{#2}
\providecommand\natexlab[1]{#1}
\providecommand\showeprint[2][]{arXiv:#2}

\bibitem[\protect\citeauthoryear{Alvarez-Picallo}{Alvarez-Picallo}{2015}]%
        {mariopicallo}
\bibfield{author}{\bibinfo{person}{Mario Alvarez-Picallo}.}
  \bibinfo{year}{2015}\natexlab{}.
\newblock \showarticletitle{MPRI Internship Report: Verification by compilation
  of higher-order functional programs}.
\newblock  (\bibinfo{year}{2015}).
\newblock


\bibitem[\protect\citeauthoryear{Appel and Palsberg}{Appel and
  Palsberg}{2002}]%
        {DBLP:books/cu/Appel2002}
\bibfield{author}{\bibinfo{person}{Andrew~W. Appel} {and} \bibinfo{person}{Jens
  Palsberg}.} \bibinfo{year}{2002}\natexlab{}.
\newblock \bibinfo{booktitle}{\emph{Modern Compiler Implementation in Java, 2nd
  edition}}.
\newblock \bibinfo{publisher}{Cambridge University Press}.
\newblock
\showISBNx{0-521-82060-X}


\bibitem[\protect\citeauthoryear{Ball, Cook, Levin, and Rajamani}{Ball
  et~al\mbox{.}}{2004}]%
        {DBLP:conf/ifm/BallCLR04}
\bibfield{author}{\bibinfo{person}{Thomas Ball}, \bibinfo{person}{Byron Cook},
  \bibinfo{person}{Vladimir Levin}, {and} \bibinfo{person}{Sriram~K.
  Rajamani}.} \bibinfo{year}{2004}\natexlab{}.
\newblock \showarticletitle{{SLAM} and Static Driver Verifier: Technology
  Transfer of Formal Methods inside Microsoft}. In
  \bibinfo{booktitle}{\emph{Integrated Formal Methods, 4th International
  Conference, {IFM} 2004, Canterbury, UK, April 4-7, 2004, Proceedings}}
  \emph{(\bibinfo{series}{Lecture Notes in Computer Science})},
  \bibfield{editor}{\bibinfo{person}{Eerke~A. Boiten}, \bibinfo{person}{John
  Derrick}, {and} \bibinfo{person}{Graeme Smith}} (Eds.),
  Vol.~\bibinfo{volume}{2999}. \bibinfo{publisher}{Springer},
  \bibinfo{pages}{1--20}.
\newblock
\showISBNx{3-540-21377-5}
\urldef\tempurl%
\url{https://doi.org/10.1007/978-3-540-24756-2\_1}
\showDOI{\tempurl}


\bibitem[\protect\citeauthoryear{Cartwright}{Cartwright}{1993}]%
        {DBLP:conf/pldi/1993}
\bibfield{editor}{\bibinfo{person}{Robert Cartwright}} (Ed.).
  \bibinfo{year}{1993}\natexlab{}.
\newblock \bibinfo{booktitle}{\emph{Proceedings of the {ACM} SIGPLAN'93
  Conference on Programming Language Design and Implementation (PLDI),
  Albuquerque, New Mexico, USA, June 23-25, 1993}}. \bibinfo{publisher}{{ACM}}.
\newblock
\showISBNx{0-89791-598-4}
\urldef\tempurl%
\url{http://dl.acm.org/citation.cfm?id=155090}
\showURL{%
\tempurl}


\bibitem[\protect\citeauthoryear{Choi, Aktemur, Yi, and Tatsuta}{Choi
  et~al\mbox{.}}{2011}]%
        {DBLP:conf/popl/ChoiAYT11}
\bibfield{author}{\bibinfo{person}{Wontae Choi}, \bibinfo{person}{Baris
  Aktemur}, \bibinfo{person}{Kwangkeun Yi}, {and} \bibinfo{person}{Makoto
  Tatsuta}.} \bibinfo{year}{2011}\natexlab{}.
\newblock \showarticletitle{Static analysis of multi-staged programs via
  unstaging translation}. In \bibinfo{booktitle}{\emph{Proceedings of the 38th
  {ACM} {SIGPLAN-SIGACT} Symposium on Principles of Programming Languages,
  {POPL} 2011, Austin, TX, USA, January 26-28, 2011}},
  \bibfield{editor}{\bibinfo{person}{Thomas Ball} {and} \bibinfo{person}{Mooly
  Sagiv}} (Eds.). \bibinfo{publisher}{{ACM}}, \bibinfo{pages}{81--92}.
\newblock
\showISBNx{978-1-4503-0490-0}
\urldef\tempurl%
\url{https://doi.org/10.1145/1926385.1926397}
\showDOI{\tempurl}


\bibitem[\protect\citeauthoryear{Clarke, Kroening, and Lerda}{Clarke
  et~al\mbox{.}}{2004}]%
        {DBLP:conf/tacas/ClarkeKL04}
\bibfield{author}{\bibinfo{person}{Edmund~M. Clarke}, \bibinfo{person}{Daniel
  Kroening}, {and} \bibinfo{person}{Flavio Lerda}.}
  \bibinfo{year}{2004}\natexlab{}.
\newblock \showarticletitle{A Tool for Checking {ANSI-C} Programs}. In
  \bibinfo{booktitle}{\emph{Tools and Algorithms for the Construction and
  Analysis of Systems, 10th International Conference, {TACAS} 2004, Held as
  Part of the Joint European Conferences on Theory and Practice of Software,
  {ETAPS} 2004, Barcelona, Spain, March 29 - April 2, 2004, Proceedings}}
  \emph{(\bibinfo{series}{Lecture Notes in Computer Science})},
  \bibfield{editor}{\bibinfo{person}{Kurt Jensen} {and}
  \bibinfo{person}{Andreas Podelski}} (Eds.), Vol.~\bibinfo{volume}{2988}.
  \bibinfo{publisher}{Springer}, \bibinfo{pages}{168--176}.
\newblock
\showISBNx{3-540-21299-X}
\urldef\tempurl%
\url{https://doi.org/10.1007/978-3-540-24730-2\_15}
\showDOI{\tempurl}


\bibitem[\protect\citeauthoryear{Cok}{Cok}{2018}]%
        {DBLP:conf/issta/Cok18}
\bibfield{author}{\bibinfo{person}{David~R. Cok}.}
  \bibinfo{year}{2018}\natexlab{}.
\newblock \showarticletitle{Reasoning about functional programming in Java and
  {C++}}. In \bibinfo{booktitle}{\emph{Companion Proceedings for the
  {ISSTA/ECOOP} 2018 Workshops, {ISSTA} 2018, Amsterdam, Netherlands, July
  16-21, 2018}}, \bibfield{editor}{\bibinfo{person}{Julian Dolby},
  \bibinfo{person}{William G.~J. Halfond}, {and} \bibinfo{person}{Ashish
  Mishra}} (Eds.). \bibinfo{publisher}{{ACM}}, \bibinfo{pages}{37--39}.
\newblock
\showISBNx{978-1-4503-5939-9}
\urldef\tempurl%
\url{https://doi.org/10.1145/3236454.3236483}
\showDOI{\tempurl}


\bibitem[\protect\citeauthoryear{Cordeiro, Kroening, and Schrammel}{Cordeiro
  et~al\mbox{.}}{2018}]%
        {DBLP:journals/sigsoft/CordeiroKS18}
\bibfield{author}{\bibinfo{person}{Lucas~C. Cordeiro}, \bibinfo{person}{Daniel
  Kroening}, {and} \bibinfo{person}{Peter Schrammel}.}
  \bibinfo{year}{2018}\natexlab{}.
\newblock \showarticletitle{Benchmarking of Java Verification Tools at the
  Software Verification Competition {(SV-COMP)}}.
\newblock \bibinfo{journal}{\emph{{ACM} {SIGSOFT} Software Engineering Notes}}
  \bibinfo{volume}{43}, \bibinfo{number}{4} (\bibinfo{year}{2018}),
  \bibinfo{pages}{56}.
\newblock
\urldef\tempurl%
\url{https://doi.org/10.1145/3282517.3282529}
\showDOI{\tempurl}


\bibitem[\protect\citeauthoryear{Federico, Payer, and Agosta}{Federico
  et~al\mbox{.}}{2017}]%
        {DBLP:conf/cc/FedericoPA17}
\bibfield{author}{\bibinfo{person}{Alessandro~Di Federico},
  \bibinfo{person}{Mathias Payer}, {and} \bibinfo{person}{Giovanni Agosta}.}
  \bibinfo{year}{2017}\natexlab{}.
\newblock \showarticletitle{rev.ng: a unified binary analysis framework to
  recover CFGs and function boundaries}. In
  \bibinfo{booktitle}{\emph{Proceedings of the 26th International Conference on
  Compiler Construction, Austin, TX, USA, February 5-6, 2017}},
  \bibfield{editor}{\bibinfo{person}{Peng Wu} {and} \bibinfo{person}{Sebastian
  Hack}} (Eds.). \bibinfo{publisher}{{ACM}}, \bibinfo{pages}{131--141}.
\newblock
\showISBNx{978-1-4503-5233-8}
\urldef\tempurl%
\url{https://doi.org/10.1145/3033019}
\showDOI{\tempurl}


\bibitem[\protect\citeauthoryear{Gaboardi, Jagannathan, Jhala, and
  Weirich}{Gaboardi et~al\mbox{.}}{2016}]%
        {DBLP:journals/dagstuhl-reports/GaboardiJJW16}
\bibfield{author}{\bibinfo{person}{Marco Gaboardi}, \bibinfo{person}{Suresh
  Jagannathan}, \bibinfo{person}{Ranjit Jhala}, {and}
  \bibinfo{person}{Stephanie Weirich}.} \bibinfo{year}{2016}\natexlab{}.
\newblock \showarticletitle{Language Based Verification Tools for Functional
  Programs (Dagstuhl Seminar 16131)}.
\newblock \bibinfo{journal}{\emph{Dagstuhl Reports}} \bibinfo{volume}{6},
  \bibinfo{number}{3} (\bibinfo{year}{2016}), \bibinfo{pages}{59--77}.
\newblock
\urldef\tempurl%
\url{https://doi.org/10.4230/DagRep.6.3.59}
\showDOI{\tempurl}


\bibitem[\protect\citeauthoryear{Jaber}{Jaber}{2018}]%
        {jabersyteci}
\bibfield{author}{\bibinfo{person}{Guilhem Jaber}.}
  \bibinfo{year}{2018}\natexlab{}.
\newblock \showarticletitle{SyTeCi: Towards automation of contextual
  equivalence for higher-order programs with references}.
\newblock  (\bibinfo{year}{2018}).
\newblock


\bibitem[\protect\citeauthoryear{Kawaguchi, Rondon, and Jhala}{Kawaguchi
  et~al\mbox{.}}{2010}]%
        {DBLP:conf/cav/KawaguchiRJ10}
\bibfield{author}{\bibinfo{person}{Ming Kawaguchi},
  \bibinfo{person}{Patrick~Maxim Rondon}, {and} \bibinfo{person}{Ranjit
  Jhala}.} \bibinfo{year}{2010}\natexlab{}.
\newblock \showarticletitle{Dsolve: Safety Verification via Liquid Types}. In
  \bibinfo{booktitle}{\emph{Computer Aided Verification, 22nd International
  Conference, {CAV} 2010, Edinburgh, UK, July 15-19, 2010. Proceedings}}
  \emph{(\bibinfo{series}{Lecture Notes in Computer Science})},
  \bibfield{editor}{\bibinfo{person}{Tayssir Touili}, \bibinfo{person}{Byron
  Cook}, {and} \bibinfo{person}{Paul~B. Jackson}} (Eds.),
  Vol.~\bibinfo{volume}{6174}. \bibinfo{publisher}{Springer},
  \bibinfo{pages}{123--126}.
\newblock
\showISBNx{978-3-642-14294-9}
\urldef\tempurl%
\url{https://doi.org/10.1007/978-3-642-14295-6\_12}
\showDOI{\tempurl}


\bibitem[\protect\citeauthoryear{Sato, Unno, and Kobayashi}{Sato
  et~al\mbox{.}}{2013}]%
        {DBLP:conf/pepm/SatoUK13}
\bibfield{author}{\bibinfo{person}{Ryosuke Sato}, \bibinfo{person}{Hiroshi
  Unno}, {and} \bibinfo{person}{Naoki Kobayashi}.}
  \bibinfo{year}{2013}\natexlab{}.
\newblock \showarticletitle{Towards a scalable software model checker for
  higher-order programs}. In \bibinfo{booktitle}{\emph{Proceedings of the {ACM}
  {SIGPLAN} 2013 Workshop on Partial Evaluation and Program Manipulation,
  {PEPM} 2013, Rome, Italy, January 21-22, 2013}},
  \bibfield{editor}{\bibinfo{person}{Elvira Albert} {and}
  \bibinfo{person}{Shin{-}Cheng Mu}} (Eds.). \bibinfo{publisher}{{ACM}},
  \bibinfo{pages}{53--62}.
\newblock
\showISBNx{978-1-4503-1842-6}
\urldef\tempurl%
\url{https://doi.org/10.1145/2426890.2426900}
\showDOI{\tempurl}


\bibitem[\protect\citeauthoryear{Vall{\'{e}}e{-}Rai, Co, Gagnon, Hendren, Lam,
  and Sundaresan}{Vall{\'{e}}e{-}Rai et~al\mbox{.}}{1999}]%
        {DBLP:conf/cascon/Vallee-RaiCGHLS99}
\bibfield{author}{\bibinfo{person}{Raja Vall{\'{e}}e{-}Rai},
  \bibinfo{person}{Phong Co}, \bibinfo{person}{Etienne Gagnon},
  \bibinfo{person}{Laurie~J. Hendren}, \bibinfo{person}{Patrick Lam}, {and}
  \bibinfo{person}{Vijay Sundaresan}.} \bibinfo{year}{1999}\natexlab{}.
\newblock \showarticletitle{Soot - a Java bytecode optimization framework}. In
  \bibinfo{booktitle}{\emph{Proceedings of the 1999 conference of the Centre
  for Advanced Studies on Collaborative Research, November 8-11, 1999,
  Mississauga, Ontario, Canada}}, \bibfield{editor}{\bibinfo{person}{Stephen~A.
  MacKay} {and} \bibinfo{person}{J.~Howard Johnson}} (Eds.).
  \bibinfo{publisher}{{IBM}}, \bibinfo{pages}{13}.
\newblock
\urldef\tempurl%
\url{https://dl.acm.org/citation.cfm?id=782008}
\showURL{%
\tempurl}


\bibitem[\protect\citeauthoryear{Wand}{Wand}{1989}]%
        {DBLP:conf/lics/Wand89}
\bibfield{author}{\bibinfo{person}{Mitchell Wand}.}
  \bibinfo{year}{1989}\natexlab{}.
\newblock \showarticletitle{Type Inference for Record Concatenation and
  Multiple Inheritance}. In \bibinfo{booktitle}{\emph{Proceedings of the Fourth
  Annual Symposium on Logic in Computer Science {(LICS} '89), Pacific Grove,
  California, USA, June 5-8, 1989}}. \bibinfo{publisher}{{IEEE} Computer
  Society}, \bibinfo{pages}{92--97}.
\newblock
\showISBNx{0-8186-1954-6}
\urldef\tempurl%
\url{https://doi.org/10.1109/LICS.1989.39162}
\showDOI{\tempurl}


\end{thebibliography}

%% Appendix
%\appendix
%\section{Appendix}
%
%Text of appendix \ldots

\end{document}